\documentstyle[12pt]{article}
\begin{document}
\vskip 2 cm
\begin{center}
\large{\bf
MATHEMATICS AS AN EXACT AND PRECISE LANGUAGE OF NATURE}
\vskip 3 cm
{\bf Afsar Abbas}\\
Institute of Physics, Bhubaneswar - 751005, India\\
{afsar@iopb.res.in}
\vskip 4 cm
{\bf ABSTRACT}
\end{center}
\vskip 2 cm

One of the outstanding problems of philosophy of science and
mathematics today is whether there is just 
"one" unique mathematics or the 
same can be bifurcated into "pure" and "applied" categories. 
A novel solution
for this problem is offered here. This will allow us to 
appreciate the manner in which mathematics 
acts as an exact and precise language of nature.
This has significant implications for Artificial Intelligence.

\newpage

Human vocal cords are capable of producing a very large ( but not
unlimited ) number of sounds of varying range in wavelength 
and frequency. As these sounds are to be distinguishable 
to the human ear, the 
number of acceptable sounds in a particular language
get still more limited. When combined together into bunches,
these provide a very large number of sound options 
available for humans to communicate with. Every language however 
chooses from amongst these large number of options to decide upon
a few acceptable ones. Thus a limited number of sound combinations
are chosen in a language to provide them with meaning and then 
the social group enforces their usage.
 
Clearly to start with a child has several options. It starts to
make various sounds, experiments with them and relishes in them.
But to survive and to be able to communicate with others it learns
that only a few sounds are acceptable to the social group that
it belongs to. So, if a child insists on calling a cat a "dog" 
and a dog a "cat" it soon learns to use the proper acceptable 
sound when it has to inform its parents as to whether it was a 
dog or a cat that bit him. A "rose" in English and "gulab"
in Hindi or Urdu are different words for the same object. But the
associations they are meant to convey are significant in those 
languages. Sound and its association is an arbitrary property of a
particular language. Clearly cultural and sociological factors
determine as to how a particular language develops.     

Note that within the periphery of a particular language a sound
which does not fall within the acceptable category is considered
gibberish. For each acceptable sound in a language there are
clearly many more gibberish sounds. Thus the range of gibberish 
sounds outside any language
is much larger than that of acceptable sounds in the language.

The above statements are supported by the following 
definitions of language. Noam Chomsky defines it thus,
" A language is a set ( finite or infinite ) of sentences each
finite in length and constructed out of a finite set 
of elements" ( Chomsky(1957) ). Also as per Trager, 
" A language is a system of arbitrary vocal symbols by means of
which the members of a society interact in terms of their 
" total culture " ( Trager(1949) ). 

Just as English is the language of residents of London and
Hindi or Urdu is of those residing in Delhi, mathematics 
is the language of scientists. And as the scientists when using
mathematics are communicating about nature,
mathematics turns out to be the language of nature.

The fact that mathematics is the language of nature has been
known to scientists for a long time. 
For example Galileo Galilie had said ( Galileo (1616) ),
" Philosophy ( ie physics) is written in this grand book  - I
mean the universe - which stands continually open to our gaze,
but it cannot be understood unless one first learns to comprehend
the language and interpret the characters in which it is written.
It is written in the language of mathematics, and its characters
are triangles, circles, and other geometrical figures, without 
which it is humanly impossible to understand a single word of it;
without these, one is wandering around in a dark labyrinth".

Bertrand Russell ( Russell (1931) ) said,
"Ordinary language is totally unsuited for expressing what physics
really asserts, since the words of everyday life are not
sufficiently abstract. Only mathematics and mathematical logic can
say as little as the physicist means to say".
James Jeans enthusiastically stated ( Jeans (1930) ),
"God is a mathematician".
And work to show that indeed mathematics is the language of
nature has been actively pursued ( Redhead (1975),
Alan and Peat(1988), French (1999), Omnes (2005) ).

It is clear that just about all the scientists and most of
the philosophers would feel that mathematics is 
indeed the language of nature. 
However the mathematics that is usable as a
language of nature is called "applied mathematics". 
Inherent in this word "applied" is the fact that there is a 
lot of mathematics which is not applied. This is the so called
"pure" mathematics. That is, mathematics which has found no
application in a description of nature. It acts outside any 
physical framework - a pure construct of human intellect as many a 
mathematician would have us believe. 
In fact, it is a dream of every mathematician to discover/invent
a mathematics which can be labelled as "pure" - that is
uncorrupted by any "lowly" application to the real world.
There must be a thrill in creating something that is absolutely
independent of any existing thing/concept/idea.
Hardy boasted ( Hardy (1940) ),
" I have never done anything ' useful '. No discovery of mine
has made, or is likely to make, directly or indirectly, for good 
or ill, the least difference to the amenity of the world. "

So clearly - though mathematics may be the language of nature 
( ie. the "applied" part ), most of it is not 
( ie the "pure" part ). What is that mathematics ? 
It seems to have a Platonic world of its own. 
The logical positivists ( Carnap ( 1995 Edition ) )
tried to understand this dichotomy by arguing that  
knowledge has two sources - the logical reasoning
and the empirical experience. 
According to them logical reasoning shall lead to
the analytical a priori knowledge.
That would embrace the field of pure mathematics. 
While the empirical experience would lead to
the synthetic a posteriori knowledge and this would correspond to
the applied mathematics. 
Explicitly or implicitly such a "dichotomous"
point of view of the intrinsic structure of mathematics is held by
most scientists, mathematicians and philosophers today. 
However, if indeed mathematics is 
the language of nature, then how come there is 
a smaller reservoir of language which 
nature communicates with ( ie applied mathematics ) but 
there is a larger reservoir of unused language ( ie pure 
mathematics )? How and why does this unused language
( " pure mathematics " ) come into existence? No one has any
reasonable understanding of it at present. This is an extremely
unsatisfactory state of affairs and demands further inquiry. 
Though not often explicitly stated, this problem remains today as
one of the most outstanding open issues in science/mathematics
and its philosophy.

If anyone has doubts as to the seriousness of the issue,
one need only read Reuben Hersh ( Hersh (2005) ).
Therein he was reviewing Ronald Omnes' new book on
philosophy of physics and mathematics (Omnes (2005)).
He quotes Omnes, " The consistency of mathematics is
therefore tantamount to the existence of
mathematically expressible laws of nature."
Thereafter Hersh goes on to say bitterly, " Minor glitches can 
be shrugged off. ( A few oddball branches of math like higher 
set theory and nonstandard logics may not be physical, 
but who cares? ) ". If this is how leading authorities
feel about the issue, then what could be more urgent?
Hersh further quotes Omnes, " What exactly is the 
extent of the present mathematical corpus that is in 
relation to the mathematics of physics? I cannot say that 
I have analyzed this question carefully, but I considered 
it from time to time when reading papers in theoretical physics
and mathematics ". 
Hence clearly up to now, no philosopher or mathematician or
scientist has understood what mathematics " really " is.  
Here I offer a novel solution to the conundrum.

Just as a child discovers the "correct" words to use 
for specific objects or ideas through social interactions,
similarly a scientist
learns the appropriate word for a particular physical reality
by interacting with nature. However, while the word "rose" 
for a particular flower is culture determined and varies from
language to language, the mathematical word for a particular
physical object is exact and specific. 
It turns out that nature is very demanding and
requires strict adherence to clear-cut mathematical rules to
reveal its reality to scientists. Only through a tortuous and
painstaking process of basically hit and trial method 
along with some judicious guesswork is it that a
scientist discovers the  " correct " mathematical 
word or expression
for a particular physical object or phenomena.

So basically, a priori, to a scientist there would exist a large
number of mathematical options/words to accurately describe a
particular physical phenomenon. 
One tries all. There is bound to be a stage where confusion
reigns. That would be the initial stage wherein
more than one mathematical model or terminology 
may appear to be applicable. 
History of science tells us that slowly with time and much
effort, the physical reality will manifest itself by demanding 
and forcing upon scientists only one particular and unique
mathematical structure. 
That will be the stage that the scientist would have
discovered the "exact" word/phrase for that particular physical 
object/phenomena. No ambiguity about that ( more on it below ). 
Thus nature has
allowed the scientist to read that particular "word".

The whole purpose of science is to continue to read nature
through this "exact" mathematical language. Acquiring a larger 
vocabulary of this mathematical language leads to a greater
fluency with nature. 

Sometimes scientists would have to develop ab initio the necessary
mathematics to understand physical reality. For example, to
understand the empirically determined Kepler's laws of planetary
motion, Newton had to develop the requisite calculus to do
so. The very fact that Newton was actually able to acquire the
necessary mathematical vocabulary made it possible for him to
appreciate the effects of gravity. 
The physical 'book' of gravity was 'read' only because
the necessary mathematics could be simultaneously developed. It
was to 'read' the other physical effects as well, that Newton's
contemporary, Leibniz was  independently developing the
required mathematics of calculus. 
Hence the requisite mathematical language of calculus 
was basic and essential to an understanding of gravity and
dynamics in physical nature. Simply put, had it not been possible
to develop the language of calculus, one would not have been
able to read nature any better. 

The basic mathematics of calculus could be developed by
scientists themselves ( ie. Newton and Leibniz ) 
as fortunately it did not necessarily require a 
too sophisticated pre-existing mathematical framework. 
Their work was simplified by the fact that the foundations of  
calculus had already been laid by earlier mathematicians. 
It was not just for nothing that
Newton had stated that he had risen on the shoulders of giants.

As there appears to be some confusion in the minds of many
as to the issue of priority here,
may I quote Richard Courant and Herbert Robbins on this matter
( Courant and Robbins (1996) p. 398 ),
" With an absurd oversimplification, the "invention" of
calculus is sometimes ascribed to two men, Newton and
Leibniz. In reality, calculus is the product 
of a long evolution that was neither initiated nor 
terminated by Newton and Leibniz,
but in which they played a decisive part."

Very often development in science is hampered if the
adequate and appropriate mathematical framework does not exist.
Therefore, had the algebra of tensors not been developed by
Einstein's contemporary mathematicians, he would never have been
able to give his equation of motion in General Theory of
Relativity in 1915. 
This equation gives the force of  gravity as an
entirely pure geometry on the left hand side of the equation while 
all the other forces - strong, weak and electromagnetic which 
describe all the matter particles and radiation, sit on the right 
hand side of the equation. This may be called the Ultimate
Equation relating space, time and matter.
This extremely beautiful and revealing 
equation describing nature could not have been 'read' but for the 
fortuitous contemporary development of tensor algebra. 

However, note that if the ideas presented here are correct ie
mathematics is indeed the language of nature, then this would
allow "anyone" ( who is sufficiently prepared ) to read it. 
And indeed in the case of the General Theory of Relativity
the German mathematician A. Hilbert simultaneous 
to Einstein, had "read" the same equation. This is being revealed
as further facts about the General Theory of Relativity are coming
to light in recent years. This is in contrast to the earlier
popular opinion that the General Theory of Relativity was the
mysterious creation of Einstein and none other. That is, as per
this opinion, had Einstein not been born there 
would have been no General
Theory of Relativity today. Misreading of history can
indeed make one appear like a blind person groping in a maze.
( This groping would be in addition to what all
scientists/mathematicians/philosophers have to do anyway
as part of their work to understand and read nature - quite a 
demanding and challenging job in itself! ).
That view was at complete variance with the fact that 
correct mathematics is indeed the language of nature and
practically anyone ( proviso sufficient mathematical and
scientific background is there ) can "read" it! 

As an attestation to the fact that mathematics
is the language of nature, history of science is replete
with examples where more than one scientist,
'read' the same language independently. 
The above example of Newton vs Leibniz and
Einstein vs Hilbert are but two such cases. 

When the quark model of particle/high energy physics was being
developed to understand the structure of the umpteen number of
particles being discovered experimentally in the 1950's and 1960's, 
a priori there were several group theoretical mathematical 
candidates for a scientific description of reality : 
the groups G2, F2, SU(2)xU(1) and SU(3) were good candidates for
it. It was basically through  
the method of hit and trial that it was found that SU(3) 
group was the correct candidate for such a description.
It was found that to understand the structure of particles like
proton, neutron etc it was necessary to assume that they were
made up of three kind of quarks which were named as up, down and
strange ( in the accepted nomenclature at present ).
As such nature 'forced' the scientists to read the word 
"SU(3)" in the quark model. Plainly stated they found that no
other mathematical 'word' can do the job appropriately!

It maybe noted here that two scientists,  
Murray Gell-Mann and G. Zweig, independently and practically
simultaneously, had come to the same 
conclusion that indeed SU(3) was the relevant
group as discussed above. So to say, they both had been able to
"read" nature correctly. Hence this is another example, in
addition to the cases of Newton/Leibniz and Einstein/Hilbert 
as pointed out above,
where proper mathematics, being the language of nature, allows
itself to be "read" by more than one person at the same time.

Another example from particle/high energy physics is that
of Quantum Chromodynamics, the theory of the strong interaction.
In the 1970's and 1980's a priori several groups SU(2), U(2),
SO(3) and SU(3) were reasonable candidates as the group
theoretical/mathematical 'words' to describe strong interaction
consistently. But experimental information and mathematical
consistency forced the group SU(3) as the relevant group for the
theory of the strong interaction to be built upon.
It has been meticulously checked and found that SU(3) and none
other is the right "word" for Quantum Chromodynamics or the
theory of strong interaction. No scientist could have even in his
wildest dreams ever thought of such a scenario right up to the
1960's.

So also is the example of the so called the Standard Model of
particle physics which is built around the group
SU(3)xSU(2)xU(1). This is the most successful model in particle
physics as of now. This mathematical structure or word is the 
result of judicious speculation, meticulous experimentation
and sheer hard work on the part of scientists all over the
world during the last 100 years or so. 
It is important to realize
that no other mathematical description can do what the Standard
Model can do. Not that scientists did not try other "words".
They did - in fact they tried very hard indeed. 
But they always failed and were forced to accept the above group. 

Another proof that indeed mathematics is the correct language of
nature is the following. It also turns out that often a particular
'word' in mathematics is used
for more than one physical situations. For example the group SU(2) 
as a mathematical language can describe the so called 'spin' 
degree of freedom and other independent 'isospin' or 'nusospin' 
degrees of freedom in physics ( Abbas(2005) ). 
Words/sounds do have independent existence. We give them a
particular meaning by association.
We match/map them to whatever physical object/concept
we wish. For example I may call a nectar 'honey' and use the
same word for my wife. As far as I am concerned the same word
'honey' is accurately describing/mapping the reality of
nectar as well as
my wife. Thus in the mathematical description of nature the group 
SU(2) can stand as the "word' for different physical entities like
spin, isospin and nusospin. 
Also as we described a little earlier the group SU(3) stood for 
quarks in the quark model as well as for another independent
framework of describing the strong interaction as the gauge force
built up around this group  - the so called Quantum
Chromodynamics. This is a further proof that
mathematics correctly read ( the SU(2) or SU(3) groups here ) 
is indeed the language of nature.

Well, good enough. However, this must be true for all that part of
mathematics which can be labelled as "applied".
But this constitutes only a small part of the whole mathematical
edifice that exists today.
What about the huge amount of "pure" mathematics.
On the basis of what has been stated so far, "pure" mathematics 
should therefore be understood so as to belong to the honourable
category of "gibberish". 
No offense meant, but as far as the language of
nature is concerned, if the relevant applied mathematics is the 
exact and accurate vocabulary of nature, then necessarily 
the 'pure' mathematics must be treated as 'gibberish' in the 
framework of what one understands as a "language". 
     
Just as a child can produce a large number of 
gibberish sounds in ordinary language so
can a mathematician produce a huge amount of 'gibberish'
mathematics. Just as the structure of the physical reality allows 
us to produce a large amount of gibberish sounds so also the
mathematical reality of nature seems to be structured in a 
manner that it allows us to conceive of a huge amount of
mathematical "gibberish". But the history of science is full of
instances of mathematics which was considered as 'pure", as
what the discoverers of the same would have us believe,
and which later turned out to be extremely valuable in physical
applications. For example as to Hardy ( who as we stated 
earlier was proud of the purity of his mathematics  ) it turned 
out that some of his work on infinite series in number theory
today finds deep applications in cryptography/communication 
theory etc. When discovered, the mathematical
matrices were thought to be beyond any
applications in nature. Today, these form the bread and butter of
quantum physicists. Hence what may be a
mathematical "gibberish" today may turn out to attain the 
status of a proper and correct word in the language to describe
nature as relevant "applied" mathematics.

If "all" of mathematics can be used to explain one or the other
aspect of nature, then there would be no "gibberish" mathematics
left.  That would mean that whatever human mind is capable of
producing in mathematics just can not go beyond 
some application or the other to nature.
It is hard to say at this stage whether this is correct. Further
work has to be done to see if this be indeed so. At this stage 
though, it appears that there is indeed a huge amount of
mathematics which finds no application in the description 
or understanding of nature. 

It should also be obvious that the way scientists learn to
read the book of nature, this should be independent of any 
cultural, sociological, historical and personal bias. The
mathematical - physical reality lies beyond our physical and
existential limitations. The example of S. Ramanujan, S N Bose
et al would clearly show that this is indeed so. 
Coming from completely different social and cultural background,
these scientists/mathematicians were still able to read the book
of nature in its exact mathematical formulation. 
In addition what they read could also be read by all the other
scientists correctly. And as such, it must be that the
mathematical mapping of physical reality,
if done in the proper manner, is accurate and exact.

Amongst those who believe that mathematics is the language of
nature there is still another issue of misunderstanding - 
and that is the issue of embedding. It is commonly believed that
" as mathematics is more fundamental and larger in
content, physics/science should be embedded in
mathematics ". For example
( French (1999) )," The relationship between mathematics and
science is clearly of fundamental concern in both the philosophy
of mathematics and the philosophy of science .... One possibility
is to employ a model - theoretic framework in which "physical
structures" are regarded as embedded in "mathematical
ones" ". Continuing in the same strain, ( Redlich (1975) ),
" it is an "empirical - historical fact" that theories in
physics can be represented as mathematical structures ".
Similar view is also expressed by Omnes ( Omnes (2005) ),
" When a physical theory .. requires mathematics in the
formalized corpus .. one can make the axioms necessary for 
the theory explicit, at least in principle, and follow the
unfolding of ideas from these axioms into 
the mathematical corpus."

As per this view, mathematics forms a basic and fundamental
structure and physics arises as a later structure which tries
to gain legitimacy by embedding itself in this already preexisting
structure. But this model leads to several interpretational 
problems. This is  necessarily artificial in content as clearly
this model is completely at variance with what has been presented
by us above. As per what we are saying here, 
in fact it is actually physics which maps the
primitive and basic reality of nature and in as much as 
mathematics is the language of nature, it continues to 'read'
this book of nature. There is no question of embedding here.
 
Many persons ( Russell (1931), Jeans (1930), Alan and Peat (1988),
Redhead (1972), Shapiro (1977), French (1999), Omnes (2005) ),
including this author,
have talked of mathematics as being the
language of science or nature. If this is so, then the ability to 
handle mathematics should be linked to the ability to handle
ordinary language grammar. However in a recent study 
Rosemary Varley et al
(2005) studied three men with brain damage which affected their
ability to handle grammar. However Varley et al found that these
men retained full ability to do computations including recursion.
They could even deal with structure - dependent concepts such
as mathematical expressions with brackets etc. This clearly 
shows that the ability to handle the language of mathematics
is entirely independent of one's ability to handle ordinary
language. Thus though both 
( human languages say English/Hindi/Urdu and
mathematics ) are languages, these are essentially different
in as much as they register differently in human brain. This
difference should be basic rather than accidental. 

Hence scientists, when making up theories have to
find judicious combinations of these disparate aspects of the two 
languages to communicate with each other. So no wonder experts in
one or the other of these two 'languages' miss appreciation of 
the total reality. Clearly this supports our contention here,
as that of mathematics being the 'exact' language of nature. 

As we have shown above, there are actually two
independent "languages". Firstly the everyday
language ( like English, Hindi and Urdu ) and secondly the
language of mathematics. The first one
is imprecise and fuzzy while the other one is precise and exact. 
It would have been rather puzzling if these two were to 
register and be controlled by the same area of the brain and in the
same manner. Because then it would have been hard to understand as
to how the same area of 
the brain could produce imprecise and fuzzy language
at the same time that it was producing another independent 
language which was precise and exact.
It shows consistency of the ideas presented here that
Varley et al found that indeed the two languages arise 
intrinsically in a different manner within the human brain.
 
If these two modes of languages register differently in the 
human brain, then it is possible that they arose at different periods
of time during human evolution, due to different requirements
of adaptation needed for them to become essential for survival.
The first, the spoken language, arose as a result of social
interaction and as a result of demands of survival for food
etc. While the second, the language of mathematics, arose as a
result of man's interaction with nature. As man spends more time
with nature than with other human beings, the second language 
must be more naturally acquired than the first one. 
This point is also supported by the fact that
other creatures have been interacting with nature 
for a longer period of time than what we humans have been
doing. Do they have a language of mathematics? Indeed they do.
When a bird needing to feed two chicks in its nest, actually
brings back two insects to feed them, then indeed it has
acquired the rudiments of mathematics.
Hence it is clear that humans in the course of evolution must 
have learnt elements of the language of nature well before
they learnt to speak. 
Therefore, it may come as a surprise to some, but 
mathematics as a language of nature, albeit in a
more elementary form,
must have been available to species other than homo sapiens.
Indeed current research shows that acquisition of
spoken language may be a much later development in human evolution. 
In fact, the growth of the human brain and the faculty of (spoken)
language acquisition may have been simultaneous ( Deacon (1992) ).

"Mistakes" are an inherent part of mathematics. Do these mistakes
occur accidentally or are they basic to mathematics itself?
Rene Descartes thought mistakes by mathematics were inadvertent.
Charles Peirce thought that these were due to lapse of memory. 
Ludwig Wittgenstein stated that actually mathematics was a
subject in which it was possible to make mistakes.
In fact in mathematics it is impossible not to make mistakes.
Riemann used what he called "Dirichlet's Principle"
incorrectly. Hilbert incorrectly thought that he had proven
the Continuum Hypothesis. There are umpteen examples.
These are only for such recorded and well known 
written cases. But when an individual scientist/mathematician,
during his private moments, in trying to go beyond known 
mathematics, keeps on making mistakes. He struggles through a maze
of mistakes and then arrives at whatever he thinks is 
consistent mathematics and is "publishable".
Only these are what we hear of and what one talks of.
These private mistakes should also be considered "mistakes"
in mathematics. These are too innumerable and are often well kept
secrets ( as never uttered ) to be recorded here! 

Just as a child when speaking a language tries to experiment with 
sounds and names, it discovers new sounds ( gibberish ) and in
fact enjoys doing it. 
That gibberish would be proven to be a
"mistake" due to social pressures and ultimately abandoned by the
child. But clearly such mistakes are part of the very process
of speech learning. So also are mistakes in mathematics.
As nature is very demanding and requires strict adherence
to its "intrinsic" mathematics so even gibberish mathematics would 
have rules of consistency. Hence though mistakes would be made
these would be subsequently corrected. In fact mistakes would be 
an inevitable part of discovering new mathematics. This analogy 
too shows that indeed mathematics is a "language" ( of nature ).

A thought, on as to how two different mathematics - pure and
applied arise. In everyday language I am free to visualize
objects which are half man - half woman or part horse - 
part dog - part snake or part cow - part human - part bird - part
elephant etc. I can visualize
groups of these as co-existing with humans. In such cases my
imagination allows me to cook up all kind of "realities" and
thus think that I am able to visualize that these may
have some kind of existence in a Platonic world of its own.
However quite clearly we
should call all these objects and their interactions etc as
nothing but gibberish, in as much as we know that these do not
correspond to any realistic objects. However it may be that
in spite of being unaware of geology and palaentology, 
that on observing a lizard at close quarters. I may
be able to imagine of a time when earth may have had extremely 
large and dangerous lizards living on it. 
That hypothesis may later on be actually proven to be correct by
palaentologists as those creatures having been
dinosaurs. So there is a small, though non-zero chance that some
of my gibberish imaginings may turn out to reflect some aspect of
reality in future.

In the same manner, mathematics being a precise and exact
language of nature - may allow us to cook up new "realities".
Imagine new mathematics - which is part this and part that,
leave an axiom here, add a lemma there, bring in some
geometry and add some algebra etc. Do I have a consistent
mathematical framework? I may make mistakes as discussed above,
but then these would be corrected in due course of time.
However,  the fact that the known
mathematics on the basis of which I am trying to go beyond,
does have consistency, hence my new mathematics
should have consistency too. If not then we abandon it.
Since different known mathematical sub-disciplines are
related to each other, hence this new mathematics may have
similar consistency and inter-relationships. 
I may call my new structure as
"pure" mathematics. However, this would be "gibberish" 
mathematics as we discussed earlier. 
It is nevertheless possible
that in future
the same structure may find applications in the description of
nature.  Hence the Platonic world of mathematics would be no more
real than the Platonic world we discussed in 
the social context above.

Phonologists have shown that phonemes in individual 
language families are quite different.
This is so because as we grow up we acquire certain pronunciation
habits that are determined by the sound patterns permitted in a
particular language ( Hjelmslev (1970) ). 

An anecdote would not be out of  order. 
A British Scientist was in Japan to attend an
International Conference. During a session, a young Japanese 
student gave a presentation of his work. The transparencies were
written in English. After the oral presentation, a senior Japanese
scientist, sitting next to him, turned to him and said, " He is
working under me. What do you think of the work?". The British
replied. " I don't really know. I would have
understood it better had he spoken in English." To which the
Senior Japanese replied, " But he was speaking in English ! ".

So though a written language may be "read" by anyone in principle,
the spoken language demands proper pronunciation which too
identifies a language. 
As Roman Jakobson has said ( Jakobson and Halle ( 1956 ) ),

" As regards the combination of linguistic elements there exists
an increasing degree of freedom. But when dealing with the
combination of distinctive features to phonemes. freedom does not
exist for the individual speaker - 
the code has already established all the possibilities that can be
realized in the given language. " 

As we have stated here, in terms of the appropriate applied 
mathematics, we are learning the proper words in the language of
nature, But as per above, how do we " pronounce " it? 
As we saw from the anecdote, proper pronunciation or lack thereof 
can provide or destroy universality in a
language. If indeed proper mathematics is the language of nature
then its " pronunciation " should be universal and exact too.

But what would one identify as " pronunciation "  in mathematics?
Here I would like to present an hypothesis as to what should be 
taken as " pronunciation " when communicating in the language
of nature.

Group theory of mathematics has been extensively applied in
physics. It it used to classify particles and in fact 
by a mathematical process called " gauging " 
these actually even define forces between particles.
A priori there are several groups as candidates to be used to
describe a particular physical phenomenon. These are for
example: infinite series of groups SO(n), SU(n) and Sp(n)
( for any n= 1,2,3,4,5, .... to infinity ) and  G2, F4. E6, E8
etc. So why was it that in the 1960's scientists discovered that
to give proper description of reality of particles only the
group SU(3) with three
quarks labelled up, down and strange was the "correct" procedure?
Why was it that it was the group combination 
SU(3) x SU(2) x U(1) that was found to be necessary in the
successful Standard Model of particle physics. 
I propose here that a
priori all these groups were options for "sound production"
to provide different "pronunciations" for that particular
"word" in the language of nature.
Nature being precise and exact chose the "pronunciation" 
as "SU(3)" for the quark model discussed above. 
And similarly it is the sound pronunciation
which is fixed in SU(3)xSU(2)xU(1). Hence as per the
suggestion here the exactness in the group 
representation is precise and exact fixing of the
sound pronunciation by nature so that unlike the spoken language,
there is no ambiguity in mathematics.

Particles have fixed quantum numbers which are used 
to identify them. For example lepton numbers for electron
and neutrino, baryon number for quarks, protons and charge number 
for electron, protons etc.
What is the nature of these quantum numbers?
In terms of what has been stated above this is just to fix 
the pronunciation to describe and identify
the different "races"/classes in the 
"genealogy"/classification of
matter in nature.

Scientists have also discovered that it is an empirical fact
that the law of gravity and that of electromagnetic forces is
inverse square of distance and not any other, say a cubic or 
a fraction or any other power of distance. This too, in 
view of what has been stated above, should be understood as 
a precise reading of the language of nature.  What the
corresponding "words" imply and how these should be interpreted
in the context of a language is 
an open issue and calls for work in future.

Within the field of Artificial Intelligence, it is important
to know as to how humans actually acquire knowledge, which is
so intrinsically related to language. 
Clearly the fact that the 
spoken human language is basically different from the
language of nature ( mathematics ), should be a 
significant fact for AI scientists. Language is more
complex than what was thought of so far.
Hence we have to redefine what we mean by intelligence in
the first place. This prompts for further work. 

In summing up, mathematics is a precise and exact language. 
As such most of the sounds/words in it 
are gibberish ( pure mathematics ) and the rest are relevant and
useful sounds ( applied mathematics ) which maps the
physical reality in an accurate manner. In the explanation
presented here there is no dichotomy of mathematics as in the
view of say the logical positivists ( and may I say that of most
of the philosophers of science as well ). In addition nature also
allows us to be able to "pronounce" the words correctly.

\newpage

{\bf REFERENCES}
\vskip 2 cm

Abbas, A. (2005), " New proton and neutron magic numbers in
neutron rich nuclei ", Modern Physics Letters, Vol A20, 2553-2560

Alan, F. and Peat, F. D. (1988), " The role of language in
science ", Foundations of Physics, vol 18, 1233-1242

Carnap, R. (1995 Edition ), " Introduction to the philosophy of
science ", Dover Publ. , New York

Chomsky, N. (1957), " Syntactic Structures ", Mouton, 
The Hague

Courant, R. and Robbins, H. (1996), " What is mathematics? ", 
Oxford University Press, Oxford

Deacon, T. W. (1992), " Brain - language coevolution ".
in " The evolution of human languages ", SFI Studies in the 
science of complexity,  Proceedings Volume X, 
Ed.: Hawkins, J. A. 
and Gell-Mann, M., Addison Wesley Publications

French, S. ( 1999) , " Models and mathematics in physics : the
role of group theory " in " From physics to philosophy ",  
Editors: Butterfield, J. and Consatantine, P., 
Cambridge University Press, Cambdridge, p. 187-207

Galileo Galilie (1616), " II Saggiatore ( The Assayer ) "

Hardy, G. H. (1940), " A mathematician's apology ", Cambridge
University Press, Cambridge

Hersh, R. (2005), " A physicist's philosophy of mathematics ",
American scientist, Vol 03 (July-August), 377-378

Hjelmskev, L. (1970),  " Language, An Introduction ", University
of Wisconsin Press, Madison

Jakobson, R. and Halle, M. (1956), " Fundamentals of language ",
Mouton, The Hague

Jeans, J. (1930), " The mysterious universe ", Cambridge 
University Press, Cambridge

Omnes, R. (2005), " Converging realities : towards a common 
philosophy of physics and mathematics ", Princeton University
Press, Princeton, U. S. A.

Redlich, M. L. G. (1975), " Symmetry in intertheory relations ",
Synthese, 32, 77-112

Russell, B. (1931), " The scientific outlook ", W. W. Norton,
New York

Shapiro, S. (1977), " Philosophy of mathematics: structure and 
ontology ". Oxford University Press, Oxford

Trager, G. (1949) , " The field of linguistics ", Battenberg 
Press, Norman OK

Varley, R. A., Klessinger, N. J. C., Romanowski, C. A. J. 
and Siegel, M., (2005), " Agrammatic but numerate ", 
Proc. Nat. Acad. Sci., vol. 102, no. 9, 3519-3524

\end{document}